# Heat Reduction by Thermal Wave Crystals

A-Li Chen*, Zheng-Yang Li, Tian-Xue Ma, Xiao-Shuang Li, Yue-Sheng Wang**

Institute of Engineering Mechanics, Beijing Jiaotong University, Beijing 100044, China

*Corresponding authors: alchen@bjtu.edu.cn

**Corresponding authors: yswang@bjtu.edu.cn

Non-Fourier heat conduction models assume wave-like behavior does exist in the heat conduction process. Based on this wave-like behavior, thermal conduction controlled in a one-dimensional periodical structure, named thermal wave crystal, has been demonstrated through both theoretical analysis and numerical simulation based on the Cattaneo-Vernotte (CV) heat-conduction model. The transfer matrix method and Bloch analysis have been applied to calculate the band structure of thermal wave propagating in thermal wave crystals. And the temperature responses are obtained by using the FDTD method, which is also used to verify the correctness of the band structure. The results show that band gaps do exist due to the Bragg scattering. Then, a calculation method to predict the mid-gap frequency of band gaps for the thermal wave crystal has been introduced in this Letter. And key parameters determining the band gaps have been discussed. This study shows the potential applications of this novel mechanism, such as thermal imagining, thermal diodes and thermal waveguides.

Traditional Fourier conduction law with implicit assumption of instantaneous thermal propagation is no longer applicable under specific conditions such as ultralow temperature, micro scale and biological tissues. In 1958, Cattaneo [1] and Vernotte [2] separately proposed a model with a time lag between the heat flux vector and the temperature gradient. In the one-dimensional (1D) case, the Cattaneo-Vernotte (CV) heat-conduction model can be written as

$$q + \tau_q \frac{\partial q}{\partial t} = -\kappa \frac{\partial T}{\partial x}, \qquad (1)$$

where $q$ and $T$ are heat flux and temperature, respectively; $\tau_q$ is the relaxation time for the phonon collision; and $\kappa$ is the thermal conductivity. The equation of energy conservation for such a problem is given by [3]

$$\frac{\partial q}{\partial x} = -\rho c_p \frac{\partial T}{\partial t} + Q, \qquad (2)$$

where $Q$ is the internal energy generation rate; $\rho$ is the density; and $c_p$ is the specific heat. Substitution of Eq. (1) into Eq. (2) yields [3]

$$\frac{1}{\tau_q}\frac{\partial T}{\partial t} + \frac{\partial^2 T}{\partial t^2} = \frac{\kappa}{\rho c_p \tau_q}\frac{\partial^2 T}{\partial x^2} + \frac{1}{\kappa}\left(Q + \tau_q \frac{\partial Q}{\partial t}\right). \qquad (3)$$

It is noted that Eq. (3) is a hyperbolic heat wave conduction equation. The heat propagates in the medium with a finite speed: $C_{CV} = \sqrt{\kappa/(\rho c_p \tau_q)}$ [3].

Due to the peculiarity of this hyperbolic wave equation, efforts have been exert on the wave-like behavior in the past several decades. Reviews of thermal propagation in the non-Fourier theory were given by Joseph *et al* [4], Tzou *et al* [5], Xu *et al* [6] among others. Wang *et al* [7] examined the non-Fourier thermal oscillation and resonance in a 1D homogeneous medium analytically with oscillatory temperature boundary conditions. Zhao *et al* [8] analyzed the non-Fourier thermal behavior in a solid sphere. Ma *et al* [9] studied the non-Fourier thermal process in functional graded materials. Furthermore, Moosaie *et al* [10-13] presented the non-Fourier effect under periodical boundary and non-periodical boundary conditions in 1D or a hollow sphere homogeneous medium analytically.

It is well known that wave manipulation is an eternal, important and challenging issue. In past decades, control of electromagnetic waves by photonic crystals [14, 15] and control of acoustic or elastic waves by phononic crystal [16-18] have been received considerable attention. We refer to Refs. [19-21] for detailed reviews in these fields. Analogously, based on the fact that thermal conduction can be modeled by ballistic phonon transport in micro-scale, control of thermal conduction by using periodic micro-scale structures has attracted considerable attention. Early, thermal conduction in superlattices was considered [22-25]. Recently, Maldovan proposed a concept of "thermocrystal" based on nano-scale phononic crystals which can manage the thermal energy flow [26-29]. Similar studies were undertaken by Zen *et al* [30], Nomura *et al* [31], Davis *et al* [32], Lacatena *et al* [33], and Anufrev *et al* [34] among others with focus on reducing the thermal conductivity by using nano-scale phononic crystals. It is noted that the aforementioned studies are limited to micro scale because the ballistic phonon transport model of thermal conduction and coherent thermal transfer is applied. Tzou [35] develop a way to relate micro-scale to macro-scale heat transfer, where the wave-like behavior (i.e. thermal wave) was included in the process of thermal transfer based on the fact of the finite time required for completing the interactions between particles. The CV model, Eqs. (1)-(3), can describe this kind of wave-like behavior. In this Letter, we will discuss thermal wave propagation through a periodically layered structure based on the CV model and Bloch theory [36]. Band gaps with pronounced heat reduction are found in the spectrum. This new class of artificial thermal material is named "thermal wave crystal" analogy to photonic and phononic crystal. It is expected to have a variety of applications such as thermal isolation, thermal diodes, thermal imaging, thermal cloaking, etc [37].

Consider a periodically layered structure with bilayer unit-cells as shown in Fig. 1. Each of the unit-cell consists of layers (sub-cells) A with thickness $l_A$ and B with thickness $l_B$ (the unit-cell's thickness $l = l_A + l_B$). All material properties $\{\kappa, \tau_q, \rho, c_p, C_{CV}\}$ of the two layers are distinguished by subscripts A and B. The coordinate (*x*, *y*) is shown in the figure. We number an arbitrary unit-cell as the *j*th unit-cell. Its left and right boundaries coordinates are $x_L^j = jl$ and $x_R^j = (j+1)l$, respectively; and the coordinates of the interface between layers A and B is $x_{AB}^j = jl + l_A$.

A 1D thermal wave propagates in the periodic structure without any internal heat source or loss (i.e. *Q*=0). For a time-harmonic thermal wave with angular frequency *ω*, the temperature and heat flux fields may be written as $\{T(x,t), q(x,t)\} = \{\hat{T}(x), \hat{q}(x)\}e^{-i\omega t}$ with $\hat{T}(x)$ satisfying

$$\hat{T}''(x) + \frac{\omega^2 + i\omega/\tau_q}{C_{CV}^2}\hat{T}(x) = 0, \qquad (4)$$

where $i = \sqrt{-1}$. The general solution of Eq. (4) is

$$\hat{T}(x) = A_1 e^{i\gamma x} + A_2 e^{-i\gamma x}, \tag{5}$$

where $A_1$ and $A_2$ are unknown coefficients, and

$$\gamma = \sqrt{\frac{\omega^2 + i\omega/\tau_q}{C_{CV}^2}}, \tag{6}$$

of which the real part demonstrates propagating of the thermal wave and the imaginary part characterizes the attenuation. The heat flux $\hat{q}(x)$ is obtained by following Eq. (2),

$$\hat{q}(x) = -A_1 \frac{i\kappa\gamma}{1-i\omega\tau_q} e^{i\gamma x} + A_2 \frac{i\kappa\gamma}{1-i\omega\tau_q} e^{-i\gamma x}. \tag{7}$$

For conciseness, the following state vector is introduced,

$$\mathbf{S}(x) = \{\hat{T}(x), \hat{q}(x)\}^{\mathrm{T}} = \mathbf{M}(x)\{A_1, A_2\}^{\mathrm{T}}, \tag{8}$$

where the superscript T denotes the transpose, and

$$\mathbf{M}(x) = \begin{pmatrix} 1 & 1 \\ -\dfrac{i\kappa\gamma}{1-i\omega\tau_q} & \dfrac{i\kappa\gamma}{1-i\omega\tau_q} \end{pmatrix} \begin{pmatrix} e^{i\gamma x} & 0 \\ 0 & e^{-i\gamma x} \end{pmatrix}. \tag{9}$$

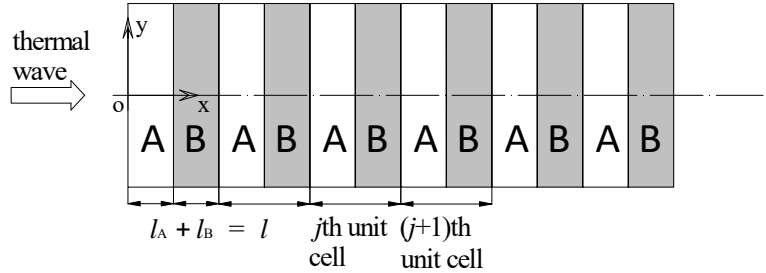

Fig. 1 Schematic diagram of a 1D thermal wave crystal.

The above solution holds for both layers A and B, and is denoted as $\mathbf{S}_A^j(x) = \mathbf{M}_A^j(x)\{A_1, A_2\}^{\mathrm{T}}$ (with $x_L^j < x < x_{AB}^j$) in layer A of the $j$th unit-cell or $\mathbf{S}_B^j(x) = \mathbf{M}_B^j(x)\{B_1, B_2\}^{\mathrm{T}}$ (with $x_{AB}^j < x < x_R^j$) in layer B. The matric $\mathbf{M}_A^j(x)$ and $\mathbf{M}_B^j(x)$ are obtained from Eq. (9) by replacing $\kappa$, $\tau_q, \gamma$, $C_{CV}$ with those with the subscripts of A and B, respectively.

Next the transfer matrix method [38, 39] is used to calculate band structures. Introduce the state vectors at the left and right boundaries of layers A and B: $\mathbf{S}_{AL}^j = \mathbf{S}_A^j(x_L^j)$, $\mathbf{S}_{AR}^j = \mathbf{S}_A^j(x_{AB}^j)$, $\mathbf{S}_{BL}^j = \mathbf{S}_B^j(x_{AB}^j)$ and $\mathbf{S}_{BR}^j = \mathbf{S}_B^j(x_R^j)$. Then from Eq. (8) it is easy to obtain the relations:

$$\mathbf{S}_{AR}^j = \mathbf{M}_{AR}^j (\mathbf{M}_{AL}^j)^{-1} \mathbf{S}_{AL}^j, \quad \mathbf{S}_{BR}^j = \mathbf{M}_{BR}^j (\mathbf{M}_{BL}^j)^{-1} \mathbf{S}_{BL}^j. \tag{10}$$

where $\mathbf{M}_{AL}^j = \mathbf{M}_A^j(x_L^j)$, $\mathbf{M}_{AR}^j = \mathbf{M}_A^j(x_{AB}^j)$, $\mathbf{M}_{BL}^j = \mathbf{M}_B^j(x_{AB}^j)$ and $\mathbf{M}_{BR}^j = \mathbf{M}_B^j(x_R^j)$.

The temperature and heat flux are continuous at the interface between two adjacent sub-layers, which states

$$\mathbf{S}_{AR}^{j} = \mathbf{S}_{BL}^{j}, \mathbf{S}_{BR}^{j} = \mathbf{S}_{AL}^{j+1}. \tag{11}$$

Combining Eq. (10) and (11), we have

$$\mathbf{S}_{AL}^{j+1} = \mathbf{M}_{BR}^{j}(\mathbf{M}_{BL}^{j})^{-1}\mathbf{M}_{AR}^{j}(\mathbf{M}_{AL}^{j})^{-1}\mathbf{S}_{AL}^{j} = \mathbf{M}_{\text{Transfer}}^{j}\mathbf{S}_{AL}^{j}, \tag{12}$$

which gives the relationship between the state vectors of the *j*th and (*j*+1)th unit cells. The matrix $\mathbf{M}_{\text{Transfer}}^{j}$ is the transfer matrix between two consecutive unit-cells, which is the same for any value of *j* and therefore is denoted as $\mathbf{M}_{\text{Transfer}}$.

To calculate the dispersion curve of a periodic structure, we generally apply the Bloch theory where the Bloch wave vector is real [36]. However, for the present thermal wave, the wave vector is complex as for the wave in a periodic viscoelastic structure [40]. Fortunately, it was proved that the Bloch theory is valid for a complex Bloch wave vector [41]. Therefore we have the following relation for the present 1D case:

$$\mathbf{S}_{AL}^{j+1} = e^{ikl}\mathbf{S}_{AL}^{j}, \tag{13}$$

where $k$ is the complex Bloch wave number and may be written as

$$k = k_{\text{Re}} + k_{\text{Im}}, \tag{14}$$

with $k_{\text{Re}}$ and $k_{\text{Im}}$ being the real and imaginary parts of $k$, respectively. Substitution of Eq. (12) into Eq. (13) yields the eigenvalue equation:

$$\mathbf{M}_{\text{Transfer}}^{j}\mathbf{S}_{AL}^{j} = e^{ikl}\mathbf{S}_{AL}^{j}, \tag{15}$$

or

$$\left|\mathbf{M}_{\text{Transfer}} - e^{ikl}\mathbf{I}\right| = 0, \tag{16}$$

where $\mathbf{I}$ is the identity matrix. Considering the detailed expression of each element of the transfer matrix $\mathbf{M}_{\text{Transfer}}$, one can obtain the following concise form of the eigenvalue equation:

$$\cosh(ikl) = \cosh(i\gamma_A l_A)\cosh(i\gamma_B l_B) + \frac{1}{2}\left(\frac{\eta_A\gamma_A}{\eta_B\gamma_B} + \frac{\eta_B\gamma_B}{\eta_A\gamma_A}\right)\sinh(i\gamma_A l_A)\sinh(i\gamma_B l_B), \tag{17}$$

where $\eta_A = \kappa_A/(1-i\omega\tau_{qA})$ and $\eta_B = \kappa_B/(1-i\omega\tau_{qB})$. The complex dispersion curves with the real frequency $\omega$ and complex wave number $k$ can be obtained by solving the complex eigenvalue equation (16) or (17).

As a numerical example, we consider a bi-layered thermal wave crystal with $l_A = l_B = 0.01$mm. The layer A is Stratum-like material, and the layer B is Dermis-like material. We refer to Ref. [42] for the material constants and list the values in Table 1.

The complex dispersion curves are illustrated in Figs. 2a and 2b which correspond respectively to the real and imaginary parts of the complex wave number. It is seen that the real part of the normalized wave number is nearly zero or 1.0 and the imaginary part is very close to zero in the frequency intervals of 0-1.8Hz, 3.5-4.5Hz and 6.7-7.9Hz. This implies that heat can transferred through the thermal wave crystal with low reduction in these frequency intervals which are therefore called "pass bands". While in the frequency intervals of 1.8-3.5Hz, 4.5-6.7Hz and 7.9-9.3Hz (the gray areas in Fig. 2b), the imaginary part has bigger values, meaning high heat reduction. These frequency ranges are called "band gaps" of which the boundaries can be determined by the extrema of the derivative of the imaginary part as showed in Fig. 2c.

TABLE 1 Material constants [42]

| Component materials | Stratum-like (Layer A) | Dermis-like (Layer B) |
|---|---|---|
| Thermal conductivity (W/(m·K)) | $\kappa_A = 0.235$ | $\kappa_B = 0.445$ |
| Specific heat (J/(kg·K)) | $c_{pA} = 3600$ | $c_{pB} = 3300$ |
| Density (kg/m³) | $\rho_A = 1500$ | $\rho_B = 1116$ |
| Relaxation time (s) | $\tau_{qA} = 1$ | $\tau_{qB} = 20$ |

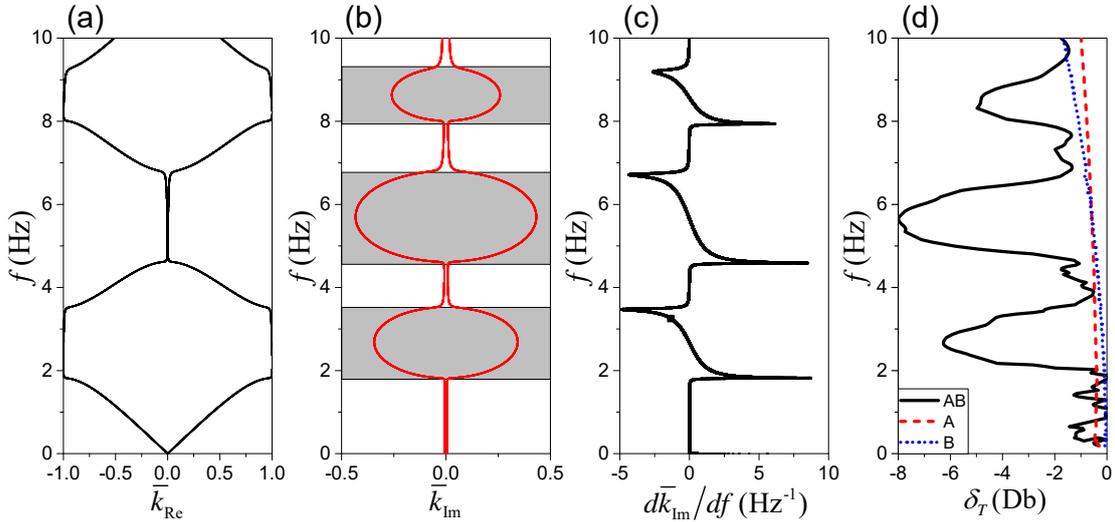

Fig. 2 Complex dispersion curves (with (a) for the real part and (b) for the imaginary part) for the thermal wave propagating in the 1D thermal wave crystal, the normalized wave number is $\bar{k} = \bar{k}_{Re} + i\bar{k}_{Im} = kl/\pi$; (c) derivative of the imaginary part of the normalized wave number shown in (b), the maxima and minima of the derivative determines the lower and upper edges of the band gaps; (d) the temperature responses $\delta_T$ calculated by the FDTD method for the thermal wave crystal of finite thickness (the black solid curve), homogeneous bulk material A (the red dashed curve) and bulk material B (the blue dotted curve), respectively.

To verify the above results from the transfer matrix method, we apply the finite difference in time domain (FDTD) method with absorbing boundary conditions (refer to Supplementary Materials [43]) to calculate the temperature response through a finite thermal wave crystal with 8 unit-cells. The ratio between the amplitudes of the oscillatory input temperature change and output temperature change ($\delta_T = \ln[\text{Amplitude}(T_{x=0})/\text{Amplitude}(T_{x=8l})]$) is plotted as the function of the frequency in Fig. 2d (see the solid curve). As comparison, the results for homogeneous bulk materials A (the dashed curve) and B (the dotted curve) with the same thickness are also shown in the figure. It can be clearly observed that the temperature decreases pronouncedly in the band gaps shown in Figs. 2a or 2b, and that the bigger values of the imaginary part of the wave number correspond to higher decrease of the temperature. However, no obvious temperature reduction is shown in homogeneous bulk materials. In other words, the thermal wave crystal exhibits more pronounced heat reduction than homogeneous materials.

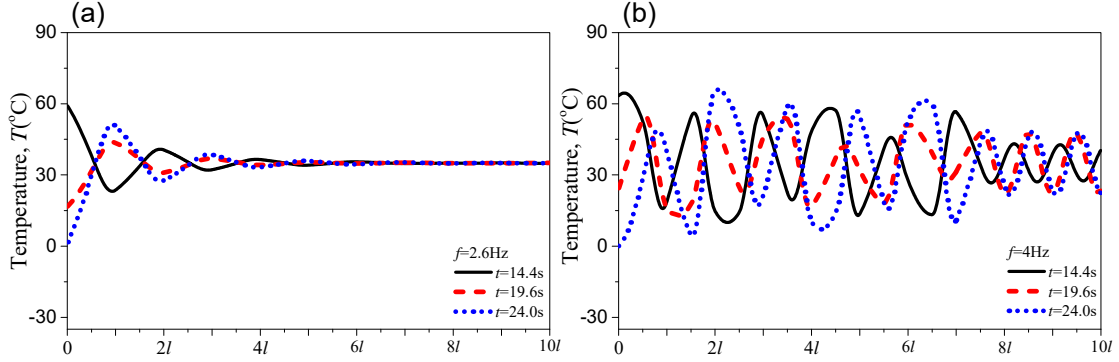

Fig. 3 The temperature distributions at different times in the thermal wave crystal of finite thickness calculated by the FDTD method: (a) $f = 2.6\,\text{Hz}$ within the first band gap, and (b) $f = 4\,\text{Hz}$ outside the band gap.

In order to understand the physics of the above phenomenon, we calculate the temperature distributions at various times in a finite thermal wave crystal of 8 unit-cells with its left end subjected to oscillatory temperature change and its right end connected with a $2l$-length strip of homogeneous medium B. The FDTD method is employed with details presented in Supplementary Materials [43]. Fig. 3 shows the temperature distributions at different times when the frequency of the oscillatory temperature change is 2.6Hz locating in the first band gap (Fig. 3a) or 4Hz outside this gap (Fig. 3b). It is seen that the change of temperature decays rapidly within the band gap (Fig. 3a) but can be transmitted through the structure outside the band gap (Fig. 3b). The wave length measured in Fig. 3a is about $2l$, which should imply that the generation of the band gaps is attributable to the Bragg-scattering like behavior as for phonons in lattice structures [16]. When thermal wave propagates in a periodical structure, multiple reflections take place at the interfaces. At certain frequencies, the trough of the reflected waves meets the crest of the origin waves which means these waves are out of phase and they interface constructively thus prevent the original wave from propagating within the thermal wave crystal [44-46]. For a Bragg-scattering type band gap, the mid-gap wave length is about $2m$ times the length of the unit-cell where $m$ is the order number of the band gap [47]. Correspondingly, the mid-gap frequency in the present layered thermal wave crystal is given by

$$f_{\text{center}}^{(m)} = \frac{m}{2(l_A/C_{VA} + l_B/C_{VB})}, \qquad (18)$$

which yields $f_{\text{center}}^{(1)} = 2.8\,\text{Hz}$ for $m=1$, $f_{\text{center}}^{(2)} = 5.6\,\text{Hz}$ for $m=2$, and $f_{\text{center}}^{(3)} = 8.4\,\text{Hz}$ for $m=3$. These values are very close to those obtained from the complex dispersion curves shown in Figs. 2a-2c (2.7Hz, 5.7Hz and 8.6Hz for the first, second and third band gaps, respectively).

To analyze the parameter dependency of the band gaps, we rewrite the dispersion equation (17) in the following nondimensional form:

$$\cosh(i\pi\bar{k}) = \cosh(in_A\bar{\gamma}_A\bar{\omega})\cosh(in_B\bar{\gamma}_B\bar{\omega}\sqrt{\bar{c}_{\rho n}/\bar{\kappa}_n})$$
$$+\frac{1}{2}\left(\frac{1}{\sqrt{\bar{c}_{\rho n}\bar{\kappa}_n}}\frac{\bar{\gamma}_B}{\bar{\gamma}_A}+\sqrt{\bar{c}_{\rho n}\bar{\kappa}_n}\frac{\bar{\gamma}_A}{\bar{\gamma}_B}\right)\sinh(in_A\bar{\gamma}_A\bar{\omega})\sinh(in_B\bar{\gamma}_B\bar{\omega}\sqrt{\bar{c}_{\rho n}/\bar{\kappa}_n}),\tag{19}$$

where $\bar{\gamma}_A = \sqrt{1+i\bar{l}/\bar{\omega}}$ and $\bar{\gamma}_B = \sqrt{\bar{\tau}_{qn}+i\bar{l}/\bar{\omega}}$; $\bar{k} = kl/\pi$ is the nondimensional wave number; $\bar{\omega} = \omega l/C_{CVA}$ is the nondimensional frequency; $\bar{l} = l/(C_{CVA}\tau_{qA})$ is the nondimensional length with $C_{CVA}\tau_{qA}$ being the mean free path; $n_A = l_A/l$ and $n_B = l_B/l$ are, respectively, the volumetric fractions of layers A and B with $n_A + n_B = 1$; $\bar{\kappa}_n = \kappa_B/\kappa_A$ is the thermal conductivity ratio; $\bar{c}_{\rho n} = \rho_B c_{pB}/(\rho_A c_{pA})$ is the volumetric thermal capacity ratio; and $\bar{\tau}_{qn} = \tau_{qB}/\tau_{qA}$ is the thermal relaxation time ratio.

Equation (20) yields nondimensional dispersion relations, $\bar{\omega}$ versus $\bar{k}$. Obviously, the key nondimensional parameters [47] affecting the band gaps are $\bar{l}$, $n_A$ (or $n_B$), $\bar{\kappa}_n$, $\bar{\tau}_{qn}$ and $\bar{c}_{\rho n}$. For the above example, the parameters are 0.959, 0.5, 1.894, 20 and 0.682, respectively. Next we check the influences of these parameters on the band gaps by changing one parameter and keeping the other four invariable. The results are illustrated in Fig. 4 where $\bar{f} = \bar{\omega}/(2\pi)$ and the color represents the value of the imaginary part of $\bar{k}$. The bright colors mean bigger values and thus the band gaps. The nondimensional mid-gap frequency, which is given by

$$\bar{f}_{center}^{(m)} = \frac{m}{2\bar{l}\left[n_A + (1-n_A)\sqrt{\bar{c}_{\rho n}\bar{\tau}_{qn}/\bar{\kappa}_n}\right]},\tag{20}$$

is also illustrated in the figure by the dashed lines.

Fig. 4a shows the effect of nondimensional length ($\bar{l}$). It is seen that the nondimensional frequency intervals do not vary with $\bar{l}$. This implies that the scaling law [49], i.e. uniformly expanding or shrinking the physical sizes of the structures by a factor $\beta$ results in the frequency spectrum being scaled by $1/\beta$, does work in this case. This result again proves Bragg scattering mechanism of the band gaps and is understood by observing Eq. (20) which is independent of $\bar{l}$.

The effect of the volumetric fraction of layer A ($n_A$) is illustrated in Fig. 4b. The band gaps first become wider and then narrower as $n_A$ increases. The upper band gaps disappear at certain medial values of $n_A$. The mid-gap frequencies increase with $n_A$, which can be interpreted by Eq. (20) (it is noted that $\sqrt{\bar{c}_{\rho n}\bar{\tau}_{qn}/\bar{\kappa}_n} > 1$ because layer A has a high thermal wave speed than layer B).

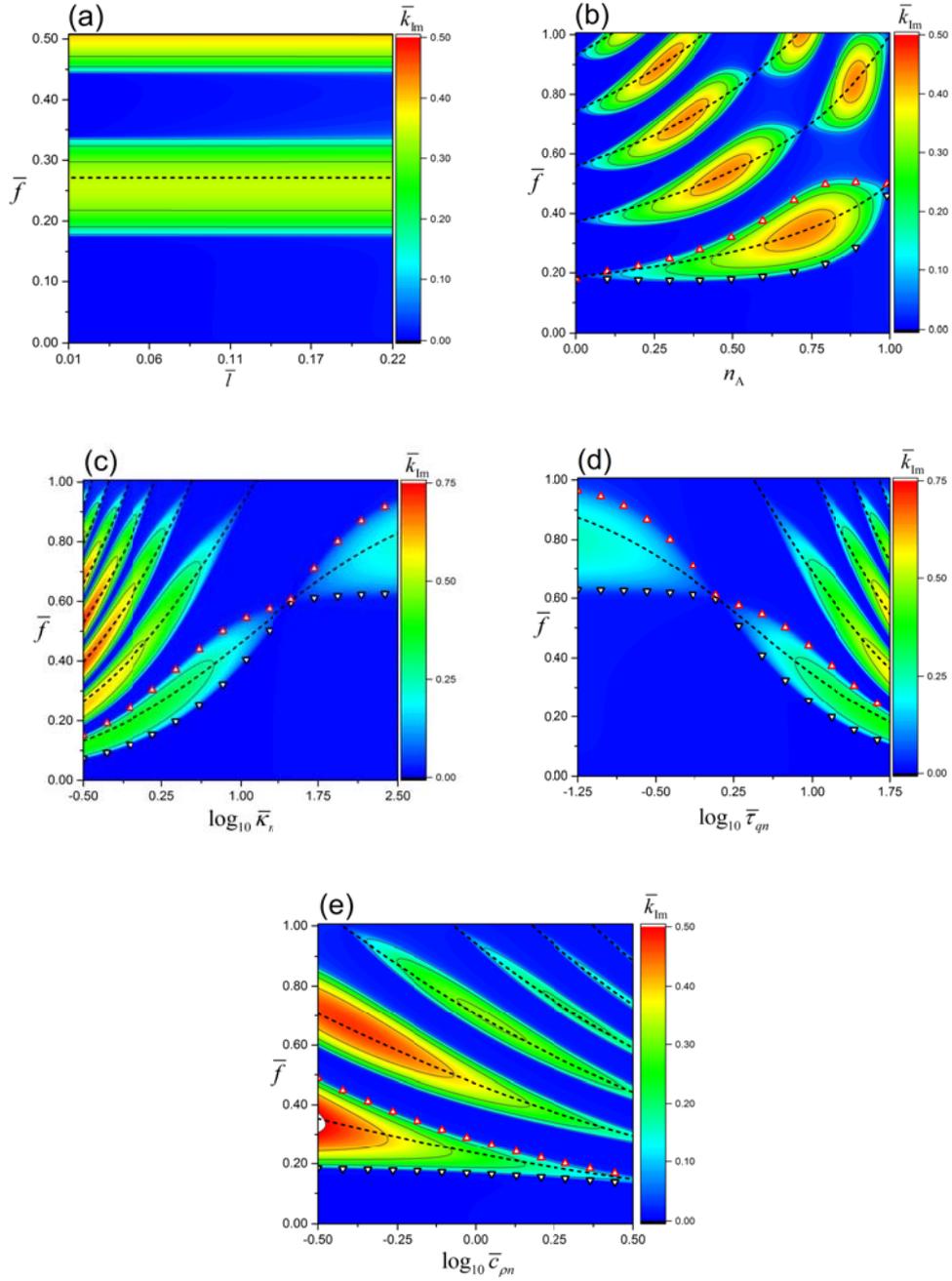

Fig .4 Effects of parameters, $\bar{l}$ (a), $n_A$ (b), $\bar{\kappa}_n$ (c), $\bar{\tau}_{qn}$ (d) and $\bar{c}_{\rho n}$ (e), on the imaginary part of wave number. The dashed lines show the mid-gap frequencies given by Eq. (20). The scattered triangles and inverted triangles denote, respectively, the upper and lower edges of the first band gap.

Fig. 4c presents the relationship of the band gaps varying with the thermal conductivity ratio ($\bar{\kappa}_n$). The band gaps first expand and then shrink with $\bar{\kappa}_n$ increasing, and disappear at certain values of $\bar{\kappa}_n$. But with continuous increase of $\bar{\kappa}_n$, the band gap appear again (see the first band gap shown in the figure). The mid-gap frequencies increase monotonically with $\bar{\kappa}_n$. Interestingly, the varying trend of the band gaps with the relaxation time ratio ($\bar{\tau}_{qn}$) shown in Fig. 4d is completely opposite that shown in Fig. 4c. This phenomenon is easily understood by considering the occurrence of $\bar{\tau}_{qn}/\bar{\kappa}_n$ in Eq. (20). Finally the band gaps changing with the volumetric heat capacity ratio ($\bar{c}_{\rho n}$) are shown in Fig. 4e. With the increase of $\bar{c}_{\rho n}$, the mid-gap frequencies decrease monotonically with the gap width decreasing for the lower band gaps but experiencing increasing-decreasing-disappearing for the upper band gaps.

In this Letter, we introduce a kind of artificial periodical structure which is named as thermal wave crystal to control heat conduction by considering the wave nature of the non-Fourier heat transfer progress described by Cattaneo-Vernotte (CV) model. The complex dispersion curves are calculated by the transfer matrix method. The transient response of the temperature field in a finite system is calculated by the FDTD method. Our results demonstrate that band gaps with pronounced heat reduction do exist in non-Fourier thermal transfer process because of the Bragg scattering. The mid-gap frequency is well predicted analytically based on the Bragg scattering mechanism. Finally the key parameters determining the band gaps are presented and discussed. We believe that the proposed artificial periodical structure may provide us a new way for thermal manipulating and heat controlling as light or sound controlling in photonic or phononic crystals. Furthermore, thermal wave crystal offers the possibility of significant new developments in thermal protection, such as the ability to safeguards biological tissues or aerospace devices during laser illuminating or thermal shock.

Financial support by the National Natural Science Foundation of China under grant number 11532001 is gratefully acknowledged.

[1] C. Cattaneo, Compt. Rendu. **247**, 431 (1958).
[2] P. Vernotte, Compt. Rendu. **246**, 3154 (1958).
[3] D. Y. Tzou, *Macro-to Microscale Heat Transfer: the Lagging Behavior* (John Wiley & Sons, Ltd. West Sussex, 2014).
[4] D. D. Joseph and L. Preziosi, Rev. Mod. Phys. **61**, 41 (1989).
[5] M. Ozisik and D. Tzou, J. Heat Transfer **116**, 526 (1994).
[6] F. Xu, K. A. Seffen, and T. J. Lu, Int. J. Heat Mass Transfer **51**, 2237 (2008).
[7] M. Xu and L. Wang, Int. J. Heat Mass Transfer **45**, 1055 (2002).
[8] W. T. Zhao, J. H. Wu, and Z. Chen, Arch. Appl. Mech. **84**, 505 (2013).
[9] X. B. Ma, F. Wang, and D. Z. Chen, J. Appl. Phys. **115**, 203505 (2014).
[10] A. Moosaie, Int. J. Heat Mass Transfer **35**, 376 (2008).
[11] A. Moosaie, Arch. Appl. Mech. **79**, 679 (2008).
[12] A. Moosaie, Int. J. Heat Mass Transfer **35**, 103 (2008).
[13] R. Shirmohammadi and A. Moosaie, Int. J. Heat Mass Transfer **36**, 827 (2009).
[14] E. Yablonovitch, Phys. Rev. Lett. **58**, 2059 (1987).
[15] S. John, Phys. Rev. Lett. **58**, 2486 (1987).
[16] M. S. Kushwaha, P. Halevi, L. Dobrzynski, and B. Djafari-Rouhani, Phys. Rev. Lett. **71**, 2022 (1993).
[17] M. Sigalas and E. Economou, J. Sound Vib. **158**, 377 (1992).
[18] M. Sigalas and E. Economou, Solid State Commun. **86**, 141 (1993).
[19] E. Kuramochi, J. Mater. Chem. C **4**, 11032 (2016).
[20] M. I. Hussein, M. J. Leamy, and M. Ruzzene, Appl. Mech. Rev. **66**, 38, 040802 (2014).
[21] V. Laude, *Phononic Crystals: Artificial Crystals for Sonic, Acoustic, and Elastic Waves* (Walter de Gruyter GmbH & Co KG, Llc. Berlin, 2015), Vol. 26.
[22] G. Chen and M. Neagu, Appl. Phys. Lett. **71**, 2761 (1997).

[23] G. Chen, Phys. Rev. B **57**, 14958 (1998).
[24] C. Dames and G. Chen, J. Appl. Phys. **95**, 682 (2004).
[25] M. N. Luckyanova *et al.*, Science **338**, 936 (2012).
[26] M. Maldovan, Phys. Rev. Lett. **110**, 025902 (2013).
[27] M. Maldovan, Nature **503**, 209 (2013).
[28] M. Maldovan, Nat. Mat. **14**, 667 (2015).
[29] A. Malhotra and M. Maldovan, J. Appl. Phys. **120**, 204305 (2016).
[30] N. Zen, T. A. Puurtinen, T. J. Isotalo, S. Chaudhuri, and I. J. Maasilta, Nat Commun **5**, 3435 (2014).
[31] M. Nomura and J. Maire, J. Elctron. Mater. **44**, 1426 (2014).
[32] B. L. Davis and M. I. Hussein, Phys. Rev. Lett. **112**, 055505 (2014).
[33] V. Lacatena, M. Haras, J. F. Robillard, S. Monfray, T. Skotnicki, and E. Dubois, Appl. Phys. Lett. **106**, 114104 (2015).
[34] R. Anufriev, J. Maire, and M. Nomura, Phys. Rev. B **93** (2016).
[35] D. Y. Tzou, J. Heat Transfer **117**, 8 (1995).
[36] L. Brillouin, *Wave Propagation in Periodic Structures: Electric Filters and Crystal Lattices* (McGraw-Hill Book Company, Inc., New York, 1953).
[37] N. Li, J. Ren, L. Wang, G. Zhang, P. Hänggi, and B. Li, Rev. Mod. Phys. **84**, 1045 (2012).
[38] R. E. Camley, B. Djafari-Rouhani, L. Dobrzynski, and A. A. Maradudin, Phys. Rev. B **27**, 7318 (1983).
[39] A. L. Chen and Y. S. Wang, Physica B **392**, 369 (2007).
[40] Y. F. Wang, Y. S. Wang, and V. Laude, Phys. Rev. B **92** (2015).
[41] F. Farzbod and M. J. Leamy, J. Vib Acoust. **133**, 051010 (2011).
[42] F. Xu, K. Seffen, and T. Lu, Int. J. Heat Mass Transfer **51**, 2237 (2008).
[43] See supplemental material for details on FDTD method, boundary conditions, intial conditons and absorbing boundarys and Refs. [50-54].
[44] S. Yang, J. Page, Z. Liu, M. Cowan, C. T. Chan, and P. Sheng, Phys. Rev. Lett. **88**, 104301 (2002).
[45] J. H. Page, S. Yang, Z. Liu, M. L. Cowan, C. T. Chan, and P. Sheng, Z. Krist-Cryst Mater **220**, 859 (2005).
[46] F. Wang, X. B. Ma, and D. Z. Chen, Thermochim. Acta **600**, 116 (2015).
[47] K. Huang and R.Q Han, *Solid State Physics* (Higher Education Press, Beijing,1996).
[48] X. Z. Zhou, Y. S. Wang, and C. Z. Zhang, J. Appl. Phys. **106**, 014903 (2009).
[49] M.S. Kushwaha, P. Halevi, G. Martinez, L. Dobrzynski, and B. Djafari-Rouhani, Phys. Rev. B **49** (1994) 2313.
[50] M. K. Zhang, B. Y. Cao, and Y. C. Guo, Int. J. Heat Mass Transfer **67**, 1072 (2013).
[51] W. K. Yeung and T. T. Lam, Numer. Heat Transfer, Part B **33**, 215 (1998).
[52] T. T. Lam and W. K. Yeung, Numer. Heat Transfer, Part B **41**, 543 (2002).
[53] R. Zhu, X. N. Liu, G. K. Hu, C. T. Sun, and G. L. Huang, Nat Commun **5**, 5510 (2014).
[54] B. Engquist and A. Majda, Proc. Natl. Acad. Sci. U.S.A. **74**, 1765 (1977).